\documentclass[12pt]{iopart}
\usepackage{graphicx}
\begin{document}
\title {Nucleation of superconductivity under rapid cycling of electric field.}
\vskip 0.5cm \author{Malay Bandyopadhyay}
\vskip 0.5cm
\address{Department of Theoretical Physics, Tata Institute of Fundamental Research, Homi Bhabha Road, Colaba, Mumbai-400005.}
\ead{malay@theory.tifr.res.in}
\begin{abstract}
\vskip 0.5cm
The effect of an externally applied high frequency oscillating electric field on the critical nucleation field of superconductivity in the bulk as well as at the surface of a superconductor is investigated in details in this work. Starting from the linearized time dependent Ginzburg-Landau (TDLG) theory and using the variational principle we have shown the analogy between a quantum harmonic oscillator with that of the nucleation of superconductivity in bulk and a quantum double oscillator with that of the nucleation at the surface of a finite sample. The effective Hamiltonian approach of Cook {\it et al} \cite{cook} is employed to incorporate the effect of an externally applied highly oscillating electric field. The critical nucleation field ratio is also calculated from the ground state energy method. The results obtained from these two approximated theories agree very well with the exact results for the case of undriven system which establishes the validity of these two approximated theories. It is observed that the highly oscillating electric field actually increases the bulk critical nucleation field ($H_{c_2}$) as well as the surface critical nucleation field ($H_{c_3}$) of superconductivity as compared to the case of  absence of electric field ($\varepsilon_0=0$). But the externally applied rapidly oscillating electric field accentuates the surface critical nucleation field more than the bulk  critical nucleation field i.e. the increase of $H_{c_3}$ is 1.6592 times larger than that of $H_{c_2}$.
\end{abstract}
\maketitle
{\section {Introduction}}
During recent years a lot of research activity is going on both in experimental and theoretical physics aiming at the understanding of dynamics of such systems which are exposed to strong time-dependent external fields \cite{manakov,chu,casati,fain}. Fundamental informations regarding high-temperature superconductors can be obtained from the high frequency electrodynamic response. Informations involving mixed state are extracted from these kind of studies \cite{gittleman,coffey,yeh,sridhar,nelson,fisher}. Also recent advances in microfabrication are creating new interesting opportunities for investigating the nucleation of superconductivity in type-II superconductor \cite{rodrigo,gielen}. If one decreases the strength of an applied magnetic field below a certain critical value, a material can become superconducting and this critical value is known as nucleation field of superconductivity. Landau and Ginzburg \cite{landau1} have shown that the value of this critical field for a bulk material ($H_{c_2}$), equals $\kappa\sqrt{2}$ ($\kappa$ is the dimensionless Ginzburg Landau parameter) times the value of its thermodynamical critical value ($H_{c}$). Saint James and de Gennes \cite{james} discovered the existence of a higher critical field, $H_{c_3}$, by considering the nucleation at the surface of a semi-infinite material. Now the main question is, whether this $H_{c_3}$ is the universal upper limit of nucleation. In other words, is the ratio $\frac{H_{c_3}}{H_{c_2}}$ a universal constant ? In this perspective, we investigate the high frequency nucleation field ratio $\frac{H_{c_3}}{H_{c_2}}$ of type-II superconductor in the present paper.\\
Time-dependent systems are generally more complicated than those of the corresponding time-independent ones. As a result, it is difficult to predict qualitative  and quantitative behavior of driven systems even in cases in which it is very easy to understand the dynamics of the corresponding time-independent ones. But there are certain methods by which these time-dependent systems can be described by an effective time-independent Hamiltonian \cite{landau2,kapitza,groz,cook,gillary,rahav}. This enables the qualitative as well as quantitative analysis of such driven systems more convincing. In order to consider the highly oscillating field we follow the sign convention of Denisov {\it et al} \cite{deni} and the references therein.\\
The Ginzburg-Landau theory for superconductivity represents one of the most useful tools available for the theoretical description of the nucleation of superconductivity in an applied field \cite{ginzburg}. It starts with a free energy expansion, completely in line with the general Landau theory  for condensed matter, with particular attention being paid on the gradient of the ordering quantity \cite{landau3}. We use the linearized time dependent Ginzburg-Landau (TDGL) theory as the starting point of my discussion about the nucleation of superconductivity \cite{kopnin,gorkov,schmid,cyrot,abrahams,caroli,saito,schmidt,tinkham}. From the linearized TDGL  theory we derive Schr$\ddot{o}$dinger like equations as that of a single harmonic oscillator and a double oscillator for the bulk nucleation of superconductivity and surface nucleation of superconductivity respectively. A. P. van Gelder have shown that the problem of nucleation resembles that of finding the ground state energy of a particle moving in a magnetic field and the ground state energy is inversely proportional to the nucleation field \cite{van}. In the present study we want to demonstrate the link between the bulk nucleation field ($H_{c_2}$) of superconductivity with that of finding the ground state energy of a single harmonic oscillator and the surface nucleation field ($H_{c_3}$) of superconductivity with that of a double oscillator. We employ the effective Hamiltonian approach \cite{landau2,kapitza,groz,cook,gillary,rahav} to incorporate the effect of highly oscillating field on the nucleation fields of superconductivity. We calculate the nucleation field ratio $\frac{H_{c_3}}{H_{c_2}}$ through the ground state energy of a single harmonic oscillator and a double oscillator for the driven as well as non-driven cases.\\
With this preceding background, we organize the rest of the paper as follows. In the next section, we discuss about the generalized linear TDGL theory of superconductivity. In this context we explore the connection between the bulk nucleation of superconductivity and the quantum harmonic oscillator and the similarity between a double oscillator and the surface nucleation of superconductivity. In section 3, we analyze the double oscillator in the presence of a high frequency field through the effective time-independent Hamiltonian method of Cook et al \cite{cook}. By calculating the ground state energy of the driven single oscillator and the driven double oscillator, we determine the nucleation field ratio, $\frac{H_{c_3}}{H_{c_2}}$, in the presence of a high frequency electric field. We summarize our findings and conclude in section 4.\\        
\section{Linearized TDGL Theory \& Nucleation of Superconductivity}
Long before the microscopic theory, a phenomenological approach to superconductivity was proposed by Ginzburg and Landau \cite{ginzburg}. The idea was that the normal-superconducting transition is a thermodynamical second order transition. So one can apply to it the general theory of second-order transitions defining an order parameter $\psi$ in such a way that $\psi$ is zero in the disordered state (normal metal) and finite in the ordered state (superconducting metal). The free energy of a superconductor is given by \cite{kopnin,tinkham}
\begin{equation}
F_{sn}=\int\Big\lbrack \alpha|\psi|^2+\frac{\lambda}{2}|\psi|^4+\frac{1}{2m^*}\Big|\Big(-i\hbar\nabla-\frac{2e}{c}\vec{A}\Big)\psi\Big|^2\Big\rbrack dv.
\end{equation}
The transition from normal to superconducting state in a magnetic field is second order and near the transition point the order parameter is small, $\psi <<\psi_{\infty}$ and hence one can easily linerize the Ginzburg-Landau equation to the following form \cite{kopnin,tinkham,degenes}
\begin{equation}
\frac{1}{2m^*}\Big(-i\hbar\nabla-\frac{2e}{c}\vec{A}\Big)^2\psi=-\alpha\psi,
\end{equation}
where $\psi$ stands for the complex superconducting order parameter and $\alpha$ is the first Ginzburg-Landau parameter, related to the temperature-dependent coherence length, $\xi(T)$, by $\alpha=-\frac{\hbar^2}{2m^*\xi^2(T)}$. The starting point of the theoretical description of nucleation ({\it i.e.} onset) of superconductivity in an applied magnetic field is this linearized Ginzburg-Landau equation (LGLE). One can easily identify that the Eq. (2) is identical with the Schr$\ddot{o}$dinger equation for a free charged particle of mass $m^*$ and charge $e^*=2e$ in a magnetic field $\vec{H}=\vec{\nabla}\times\vec{A}$, with $-\alpha=|\alpha|$ playing the role of the energy eigenvalue. This property allows us to apply various familiar solutions and methods of usual quantum mechanics to the problem of nucleation in superconductivity. The lowest eigenvalue of the LGLE gives the highest magnetic field at which the nucleation of the superconductivity can occur. Now the big question is what happens in nonstationary cases ? For instance, if one applies an external electric field which varies very slowly, is it possible that the order parameter will be given by the same equation as in the static case where the time is a parameter. On the other hand, if the field varies very rapidly with time will it be possible that the superconductor will respond to an average of the field as it happens in other systems in physics ? The last quary is our basic investigation in this work. Time dependent Ginzburg Landau (TDGL) model often gives a reasonable picture of superconducting dynamics \cite{abrahams,caroli}. Unlike from its static counterpart, validity of the TDGL theory is much more limited. It is not enough just to be close to the critical temperature. The necessary condition is that the deviation from equilibrium is small; the quasiparticle excitations should remain essentially in equilibrium with the heat bath. It can normally be fulfilled for gapless superconductor \cite{kopnin, cyrot}. So, we begin by writing down the time-dependent Ginzburg Landau  equation that governs the dynamics of the superconducting order parameter \cite{abrahams,caroli} :
\begin{equation}
i\hbar\Big(\frac{\partial \psi}{\partial t}+\frac{2ie\phi}{\hbar}\Big)=-|\alpha|\psi+\lambda|\psi|^2\psi+\frac{1}{2m^*}\Big|\Big(-i\hbar\nabla-\frac{2e}{c}\vec{A}\Big)\Big|^2\psi.
\end{equation}
Now choosing the electrical potential $\phi=-\frac{\varepsilon_0}{2}x\cos(\omega t)$, $H$ along the z axis with convenient gauge $A_y=Hx$ and linearizing one can show
\begin{equation}
\Big\lbrack-\frac{\hbar^2}{2m^*}\nabla^2+\frac{i\hbar e}{m^*c}H x \frac{\partial}{\partial y}+\Big(\frac{2e^2H^2}{m^*c^2}\Big)x^2\Big\rbrack\psi-e\varepsilon_0x\cos(\omega t)\psi= i\hbar\frac{\partial\psi}{\partial t}+|\alpha|\psi.
\end{equation}
Now using $\psi = e^{ik_zz}e^{ik_yy}f(x)e^{-i\frac{e\varepsilon_0x\sin(\omega t)}{\hbar \omega}}$ and then following the method of Cook {\it et al} \cite{cook} one obtains
\begin{equation}
-\frac{\hbar^2}{2m^*}\frac{d^2f}{dx^2}+\frac{2e^2H^2}{m^*c^2}(x-x_0)^2f=\Big(|\alpha|+\frac{e^2\varepsilon_0^2}{4m^*\omega^2}-\frac{\hbar^2k_z^2}{2m^*}\Big)f,
\end{equation}
where $x_0=\frac{\hbar k_y}{2eH}$.\\
\subsection{Bulk Superconductivity, Single Harmonic Oscillator and $H_{c_2}$}
In this subsection, we calculate the critical nucleation field of superconductivity in the presence of an external magnetic field for a large sample. We consider the large sample in the presence of a magnetic field $\vec{H}$ along the $z$ axis and a convenient gauge is $A_y=Hx$. Equation (5) is our starting point of discussion about bulk nucleation. Equation (5) is same as that of a Schr$\ddot{o}$dinger equation for a particle of mass $m^*$ bound in a harmonic oscillator potential with force constant $\frac{4e^2H^2}{m^*c^2}$. The resulting harmonic oscillator eigenvalues are
\begin{equation}
E_n=\Big(n+\frac{1}{2}\Big)\hbar\omega_0=\Big(n+\frac{1}{2}\Big)\hbar\Big(\frac{2eH}{m^*c}\Big).
\end{equation}
In view of Eq. (5) these energy eigenvalues, $E_n$, are to be equated to $\Big(|\alpha|+\frac{e^2\varepsilon_0^2}{4m^*\omega^2}-\frac{\hbar^2k_z^2}{2m^*}\Big)$. Thus
\begin{equation}
H=\frac{m^*c}{(2n+1)e\hbar}\Big(|\alpha|+\frac{e^2\varepsilon_0^2}{4m^*\omega^2}-\frac{\hbar^2k_z^2}{2m^*}\Big).
\end{equation}
Here, we are concerned about the highest value of $H$ {\it i.e.} $H_{c_2}$ which is obviously given by the lowest eigenvalue ($n=0$, $\varepsilon_0=0$ and $k_z=0$). Thus 
\begin{equation}
H_{c_2}^{old}=\frac{m^*c|\alpha|}{e\hbar}=\frac{\phi_0}{2\pi\xi^2}=\kappa\sqrt{2}H_{c},
\end{equation}
where flux quantum $\phi_0=\frac{hc}{2e}$. The above relation $H_{c_2}^{old}=\kappa\sqrt{2}H_c$ gives us an important message. When $\kappa>\frac{1}{\sqrt{2}}$, $H_{c_2}>H_c$ and vortex phase in the type-II superconductor appears. On the other hand for $\kappa<\frac{1}{\sqrt{2}}$, $H_{c_2}<H_c$ and Meissner effect sets in, the mixed phase does not appear and one obtains type-I superconductor.\\
On the other hand, for $\varepsilon_0\neq 0$,
\begin{equation}
H_{c_2}^{new}=\frac{m^*c(|\alpha|+\frac{e^2\varepsilon_0^2}{4m^*\omega^2})}{e\hbar}=\kappa\sqrt{2}H_c+\frac{c}{e\hbar}\frac{e^2\varepsilon_0^2}{4\omega^2}=H_{c_2}^{old}+\frac{ec\varepsilon_0^2}{4\hbar\omega^2}.
\end{equation}
It is evident from Eq. (9) that the highly oscillating electric field actually increases the nucleation field by an amount $\frac{ec\varepsilon_0^2}{4\hbar\omega^2}$ in the bulk nucleation of superconductivity.
\subsection{Surface Superconductivity, Double Oscillator and $H_{c_3}$}
So far in my treatment of the Ginzburg-Landau equation at the mean-field level we have not taken into account the surface of the sample. At the surface of the superconductor, some additional boundary conditions need to be imposed on the solutions. One can quite reasonably expect that the presence of an interface between the superconductor and a non-superconducting material, such as a normal metal or an insulator must affect the nucleation of superconductivity in the material. We consider a specimen with a single plane surface and the external magnetic field to be parallel to the surface {\it i.e.} $\vec{H}=H\hat{z}$. The superconducting sample is located in the half-space $x>0$, while we take the non-superconducting material to be located in the half space $x<0$. The latter material is taken to be either vacuum or an insulating material.
\begin{figure}[h]
\begin{center}
{\rotatebox{0}{\resizebox{6cm}{4cm}{\includegraphics{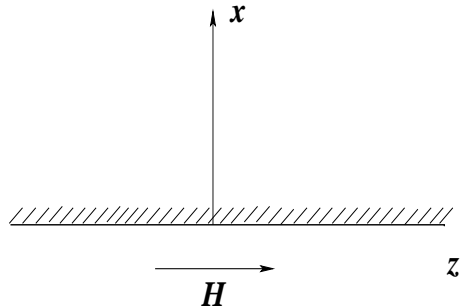}}}}
\caption{The finite sample in the upper half space $x>0$.}
\end{center}
\end{figure}
Then the superconducting boundary condition imposed on $\psi$ in finite samples \cite{degenes}:
\begin{equation}
\Big(-i\hbar\nabla-\frac{2e}{c}\vec{A}\Big)\psi\Big|_{n}=0.
\end{equation} 
It reduces to the Neumann boundary condition, $\nabla\psi|_n=0$, when the magnetic vector potential, $\vec{A}$, can be chosen in a form with zero normal component at the boundary of the sample. In my case this becomes 
\begin{equation}
\frac{\partial \psi}{\partial x}\Big{|}_{x=0}=0,
\end{equation}
where $\vec{A}=Hx\hat{y}$. We look for a solution of the form $\psi=e^{ik_yy}e^{-i\frac{e\varepsilon_0x\sin(\omega t)}{\hbar \omega}}f(x)$ for the linearized TDGL equation (Eq. 4) with the constraints :
\begin{eqnarray}
\frac{df}{dx}\Big|_{x=0}=0\\
\frac{df}{dx}\Big|_{x=\infty}=0.
\end{eqnarray}
The complication arises because the boundary condition states that the solution must be flat at a position $x=0$ while the minimum is located at $x=x_0$. When the minimum of the potential is located far from the surface ($x_0=\infty$) one can ignore the boundary condition, and when $x_0=0$ the boundary condition is satisfied by the standard solution of usual Schr$\ddot{o}$dinger equation of a harmonic oscillator. Thus in both the cases we obtain $H=H_{c_2}$. One can easily understand that the surfaces have consequences for the solution to the LGLE only at the intermediate values of $x_0$, {\it i.e.} $0<x_0<\infty$. For the intermediate values of $x_0$, we can think of solving the Schr$\ddot{o}$dinger equation by employing the effective Hamiltonian method of Cook {\it et al} \cite{cook} and thus the effective Schr$\ddot{o}$dinger like equation becomes
\begin{equation}
\frac{-\hbar^2}{2m^*}\frac{d^2f}{dx^2}+\frac{2e^2H^2}{m^*c^2}\Big(x - x_0)^2f- \frac{e^2\varepsilon_0^2}{4m^*\omega^2}f=-\alpha(x_0)f.
\end{equation}
This is an eigenvalue problem where the eigenvalue itself is $x_0$ dependent, and my task is to minimize this with respect to $x_0$ subject to the boundary condition on $f$ at the surface. Introducing $\chi=\sqrt{\frac{m^*\omega_0}{\hbar}}x$, $\chi_0=\sqrt{\frac{m^*\omega_0}{\hbar}}x_0$, $\Omega=\sqrt{\frac{m^*\hbar\omega_0}{2}}\omega$ and $\beta =-\frac{2\alpha}{\hbar\omega_0}$, one can rewrite Eq. (14) as follows :
\begin{equation}
-\frac{d^2f}{d\chi^2}+(\chi-\chi_0)^2f-\frac{e^2\varepsilon_0^2}{4\Omega^2}f =\beta f.
\end{equation}
Now my task is to find the lowest possible value of $\beta$ subjected to the boundary conditions $\frac{df}{d\chi}=0$ at $\chi=(0,\infty)$. This can be phrased as the following variational problem of minimizing the functional
\begin{equation}
\beta=\frac{\int_0^{\infty}d\chi\Big\lbrack\Big(\frac{df}{d\chi}\Big)^2+(\chi-\chi_0)^2f^2-\frac{e^2\varepsilon_0^2}{4\Omega^2}f^2\Big\rbrack}{\int_0^{\infty}d\chi f^2}
\end{equation}
with respect to variations in $f$. The Euler-Lagrange equation for this variational problem is precisely the scaled differential Eq. (15). To do the minimization, we use the following trial wave-function :
\begin{equation}
f(\chi)=\exp\Big\lbrack-\frac{1}{2}b\chi^2\Big\rbrack.
\end{equation}
With the help of this trial wave-function, we obtain
\begin{eqnarray}
\beta&=&\frac{\int_0^{\infty}d\chi\lbrace(-b\chi)^2+(\chi^2-2\chi\chi_0+\chi_0^2)-\frac{e^2\varepsilon_0^2}{4\Omega^2}\rbrace\exp(-b\chi^2)}{\int_0^{\infty}d\chi\exp(-b\chi^2)}\nonumber \\
&=&\frac{\sqrt{\frac{\pi}{b}}\Big(\frac{b}{4}+\frac{1}{4b}+\frac{\chi_0^2}{2}\Big)-\frac{\chi_0}{b}-\frac{e^2\varepsilon_0^2}{8\Omega^2}\sqrt{\frac{\pi}{b}}}{\frac{1}{2}\sqrt{\frac{\pi}{b}}}\nonumber \\
&=&\frac{b}{2}+\frac{1}{2b}+\chi_0^2-\frac{2\chi_0}{\sqrt{\pi}b}-\frac{e^2\varepsilon_0^2}{4\Omega^2}.
\end{eqnarray}
Now  minimizing $\beta$ with respect to $\chi_0$, we obtain
\begin{equation}
\frac{\partial \beta}{\partial \chi_0}=-\frac{2}{\sqrt{\pi b}}+2\chi_0=0
\end{equation}
which yields $\chi_0=\frac{1}{\sqrt{\pi b}}$. Substituting this back in $\beta$ and again minimizing with respect to b, we find
\begin{equation}
\frac{\partial \beta}{\partial b}=\frac{1}{2}-\frac{1}{2b^2}+\frac{1}{\pi b^2}=0
\end{equation}
and it gives us $b=\sqrt{1-\frac{2}{\pi}}$. Substituting this back into $\beta$, we obtain
\begin{equation}
\beta_{min}=\sqrt{1-\frac{2}{\pi}}-\frac{e^2\varepsilon_0^2}{4\Omega^2}.
\end{equation}
From the definition of $\beta$ one can relate 
\begin{equation}
\beta_{min}=-\frac{2\alpha}{\hbar\omega_0}=\frac{\hbar c}{2eH\xi^2}=\frac{H_{c_2}^{old}}{H}.
\end{equation}
Thus
\begin{equation}
H=\frac{H_{c_2}^{old}}{\sqrt{1-\frac{2}{\pi}}-\frac{e^2\varepsilon_0^2}{4\Omega^2}}=\frac{H_{c_2}^{old}}{\sqrt{1-\frac{2}{\pi}}-\frac{ec\varepsilon_0^2}{4\hbar\omega^2H}}.
\end{equation}
In view of equation (23) one can realize that the physically accessible region is defined as follows : $\sqrt{1-\frac{2}{\pi}}>\frac{e^2\varepsilon_0^2}{4\Omega^2}>0$. 
Now solving equation (23) for `H' one obtains
\begin{equation}
H=H_{c_3}^{new}=\frac{H_{c_2}^{old}+\frac{ec\varepsilon_0^2}{4\hbar\omega^2}}{\sqrt{1-\frac{2}{\pi}}}=\frac{H_{c_2}^{new}}{\sqrt{1-\frac{2}{\pi}}}=H_{c_3}^{old}+\frac{ec\varepsilon_0^2}{4\hbar\omega^2\sqrt{1-\frac{2}{\pi}}}.
\end{equation}
From Eq. (24), we obtain for $\varepsilon_0=0$, $H_{c_3}^{old}=1.6592H_{c_2}^{old}$ and for $\varepsilon_0\neq 0$, the surface nucleation field $H_{c_3}^{new}=1.6592H_{c_2}^{new}$. The main message of Eq. (24) is that when the superconductivity occurs in a system, it starts to nucleate at the surface of an ideal defect-free sample and not in the interior of the sample. If the sample has defects in the interior, superconductivity starts to nucleate in the vicinity of such defects. On the other hand the high frequency field further accentuates the nucleation at the surface rather than in the interior of the sample which is evident from equation (24). For the bulk sample the increasing amount due to the rapidly oscillating field is $\frac{ec\varepsilon_0^2}{4\hbar\omega^2}$ (see equation 9). On the other hand the rapidly oscillating field increases the surface nucleation field by an amount $1.6592\frac{ec\varepsilon_0^2}{4\hbar\omega^2}$ (see equation 24). Thus the rapidly oscillating electric field accentuates the surface nucleation field by $1.6592$ times more than the bulk nucleasion field increasing factor. From equation (9) and equation (24) one can determine the real values for this enhancement in nucleation field. If an experimentalist uses $\varepsilon_0=0.01$ V/m, $\omega=10$ GHz, the enhancement of the bulk nucleation field is $0.114$ T more than the without driven result. For the same values of electric field strength and frequency, the enhancement of surface nucleation field is $0.189$ T more than the non-driven system. \\
Before concluding this subsection we want to show the analogy between the problem of nucleation of superconductivity at the surface with that of a double oscillator. Equation (14) is still have the form of Schr$\ddot{o}$dinger equation for a particle in a harmonic well centered at $x_0$, but the boundary condition, Eqs. (12) and (13), means that the eigenvalue depends crucially on the value of $x_0$.
\begin{figure}[h]
\begin{center}
{\rotatebox{270}{\resizebox{8cm}{8cm}{\includegraphics{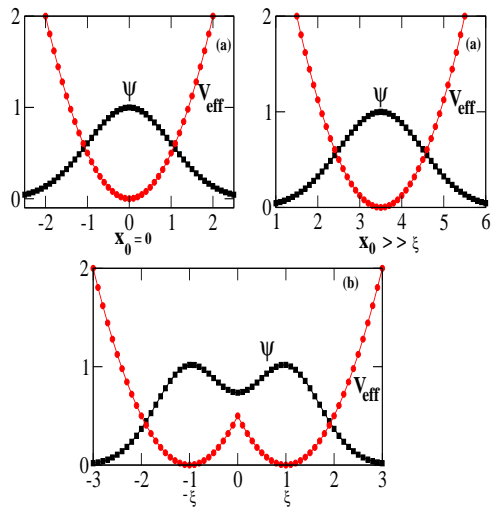}}}}
\caption{(color online) (a) Surface and interior nucleation at $H_{c_2}$ with single oscillator potential form; (b) Surface nucleation at $H_{c_3}$ with double-oscillator potential form.}
\end{center}
\end{figure}
One can incorporate the boundary condition by an image method. We consider a particle moving in the potential $V(x)$ of Fig. 2(b) which is my original harmonic well together with its reflection at the surface. The ground-state wave function $\psi_0(x)$ in $V(x)$ must be symmetric about $x=0$, as required in Eq. (12), and for $x>0$ it satisfies Eq. (14). Thus $\psi_0(x)$ for $x>0$ is the solution to our problem, and the corresponding eigenvalue $E_0$ gives the critical field. We can compare $E_0$ for various $x_0$ with the eigenvalue $E$ in the single harmonic well. For $x_0\rightarrow\infty$, $E_0\rightarrow E$, and for $x_0=0$, $E_0=E$ again. On the other hand for intermediate values of $x_0$, $E_0$ is less than $E$ because $V(x)$ is smaller than the single harmonic potential in some region. The new surface eigenfunction must have a lower eigenvalue than the interior ones because it arises from a potential that is lower and broader than the single harmonic well about $x_0$. Thus one can say that the problem of nucleation of superconductivity at the surface is analogous to that of double oscillator problem.\\
\section{Driven Double and Single Oscillator : Critical Field Ratio}
The analysis of the previous section has clearly demonstrated that the problem of nucleation of superconductivity in the bulk and at the surface are analogous to that of a single harmonic oscillator and double oscillator respectively. A. P. van Gelder had shown that the problem of nucleation resembles that of finding the ground state energy of a quantum particle and the nucleation field is inversely proportional to the ground state energy \cite{van}. Henceforth, one can easily calculate the nucleation field ratio, $\frac{H_{c_3}}{H_{c_2}}$, by calculating the ground state energies of a single harmonic oscillator and  a double oscillator. Due to the recent developments in the regime of superconducting high frequency devices \cite{asle,winkler}, we would like to exhibit the effect of high frequency field on the nucleation of superconductivity in this section. The effect of high frequency field can be taken care of by simply following the effective Hamiltonian method \cite{landau2,kapitza,groz,cook,gillary,rahav}. By following Cook {\it et al} \cite{cook}, we convert the time-dependent problem into an effective time independent one and then calculate the ground state energy by following standard stationary quantum mechanics procedures. Thus the main objective of this section is to find the nucleation field ratio, $\frac{H_{c_3}}{H_{c_2}}$, from the eigenvalue solutions of the single harmonic oscillator and the double oscillator in both the cases {\it i.e.} in the presence and absence of the high frequency field.\\    
We consider a quantum particle  moving in a double oscillator potential $V_0(|x|)=\frac{1}{2}m^*\omega_0^2\Big(|x|-x_0\Big)^2$ and  driven by a high frequency monochromatic force. Thus the driven double-oscillator Hamiltonian becomes
\begin{equation}
{\cal{H}}=\frac{p^2}{2m^*}+\frac{1}{2}m^*\omega_0^2\Big(|x|-x_0\Big)^2-ax\cos(\omega t).
\end{equation}
Now following Cook {\it et al} \cite{cook} we can express the effective time-independent potential 
\begin{equation}
V_{eff}(x)= \frac{1}{2}m^*\omega_0^2\Big(|x|-x_0\Big)^2-\frac{a^2}{4m^*\omega^2}.
\end{equation}
To determine ground state energy of this driven double oscillator, we consider the normalized ground-state eigenfunctions of single oscillators centered at $\pm{x_0}$ :
\begin{equation}
u_{\pm}=\Big(\frac{\eta}{\sqrt{\pi}}\Big)^{\frac{1}{2}}e^{-(\gamma\mp{\gamma}_0)^2},
\end{equation}
where we have introduced dimensionless variables $\gamma=\eta x$ and $\gamma_0=\eta x_0$
 with $\eta=\sqrt{\frac{m\omega_0}{\hbar}}$. Now, we compute the expectation value of the Hamiltonian $\langle{\cal{H}}\rangle$, in the state $\psi_0=\frac{1}{\sqrt2}(u_++u_-)$. The norm of $\psi_0$ is $N_0=\langle\psi_0|\psi_0\rangle=1+e^{-{\gamma}_0^2}$. To compute the expectation value of the kinetic energy operator ($\hat{T}=-\frac{\hbar^2}{2m^*}\frac{d^2}{dx^2}$) we need to use the identity 
\begin{equation}
\hat{T}u_{\pm}=\frac{\eta^2}{2}\lbrack u_{\pm}-(\gamma\mp{\gamma}_0)^2u_{\pm}\rbrack.
\end{equation} 
Thus, we find
\begin{equation}
\langle \hat{T} \rangle_0 = \frac{\langle \psi_0|\hat{T}|\psi_0\rangle}{\langle\psi_0|\psi_0\rangle} = \frac{\eta^2\hbar^2}{4m^*}\Big(1-\frac{2{\gamma}_0^2}{e^{{\gamma}_0^2}+1}\Big).
\end{equation}
To obtain the expectation value of the potential energy operator $\hat{V}=\frac{1}{2}m^*\omega_0^2\Big(x^2-2{x}_0|x|+{x}_0^2\Big)-\frac{a^2}{4m^*\omega^2}$, we need to compute $\langle\psi_0|x^2|\psi_0\rangle$ and $\langle\psi_0||x||\psi_0\rangle$. So, we obtain
\begin{eqnarray}
\hskip-1.8cm
\langle V \rangle_0&=&\frac{1}{2}m^*\omega_0^2\Big\lbrack \frac{1}{2\eta^2}+\frac{{\gamma}_0^2}{(1+e^{-{\gamma}_0^2})\eta^2}-\frac{2{x}_0}{(1+e^{-{\gamma}_0^2})\eta}\Big({\gamma}_0 \rm{erf}({\gamma}_0)+\frac{2}{\sqrt{\pi}}e^{-{\gamma}_0^2}\Big)+{x}_0^2\Big\rbrack\nonumber \\
\hskip-1.8cm
&&-\frac{a^2}{4m^*\omega^2}.
\end{eqnarray} 
Finally the ground state energy of the driven oscillator is given by
\begin{equation}
\langle {\cal {H}}\rangle_0 = \frac{E_0}{\hbar\omega_0} = \frac{1}{2}+\frac{{\gamma}_0^2}{1+e^{-{\gamma}_0^2}}\Big(1- \rm{erf}({\gamma}_0)-\frac{2}{\sqrt{\pi}{\gamma}_0}e^{-{\gamma}_0^2}\Big)-\frac{a^2}{4m^*\hbar\omega_0\omega^2}.
\end{equation}
\begin{figure}[h]
\begin{center}
{\rotatebox{270}{\resizebox{8cm}{8cm}{\includegraphics{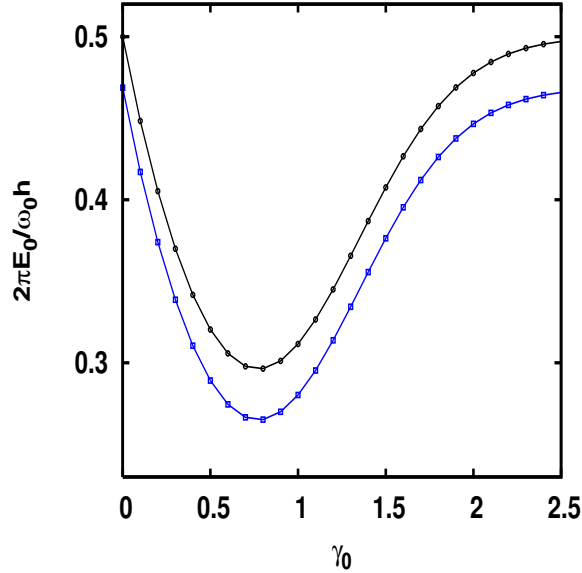}}}}
\caption{(color online) Plot of ground state energy versus ${\gamma}_0$ for the double oscillator in the presence and absence of the rapidly oscillating electric field with $a=1.0$, $\omega=4.0$, $\omega_0=0.5$ (blue filled square) and $a=0$ (black filled circle) respectively.}
\end{center}
\end{figure}
In Fig. 3, we plot the ground-state energy versus ${\gamma}_0$ for the  double-oscillator in the presence as well as absence of the external high frequency field. As pointed out by A. P. van Gelder the critical nucleation field is inversely proportional to the ground state energy. Thus, we calculate the critical nucleation field ratio ($\frac{H_{c_3}}{H_{c_2}}$) from the numerical computation of the minimum of ground state energies of the double and single oscillators (Eq. (31)). Minimum ground state energies for the double oscillator (dosc) in the presence and absence of the external field are computed from Eq. (31) by putting $a=1.0$, $\omega=4.0$, $\omega_0=0.5$ and $a=0$ respectively. On the other hand the ground state energies for the single oscillator (sosc.) in the presence and absence of the field  ($a=0$) can be computed from the Eq. (31) by simply putting $\gamma_0=0$. From our computation, we obtain the minimum of ground state energies for the double-oscillator with external rapidly oscillating force (wf) and without force (wof) cases are $\langle H \rangle_{0,min}^{wf}=0.2808$ and $\langle H \rangle_{0,min}^{wof}=0.2964$ respectively. Thus the ratio of the critical fields for the force-free case is given by
\begin{equation}
\frac{H_{c_3}^{old}}{H_{c_2}^{old}}=\frac{\langle H \rangle_{0,sosc.}^{wof}}{\langle H \rangle_{0,dosc.}^{wof}}=\frac{0.5}{0.2964}=1.6869.
\end{equation}
We have to mention that a more careful numerical analysis based on the hypergeometric functions in the non-driven system yields (exact result)
\begin{equation}
\frac{H_{c_3}^{old}}{H_{c_2}^{old}}=1.695.
\end{equation}
From the variational principle one obtains $H_{c_3}^{old}=1.6592 H_{c_2}^{old}$ and from the ground state energy method, we obtain $H_{c_3}^{old}=1.6869 H_{c_2}^{old}$. Now comparing these two results with equation (33) one can say that the results obtained from the two approximated theories (variational principle and ground state enrgy) agree very well with the exact results. This establishes the validity of these two approximated theories.\\
Now, we employ the ground state energy method to deternine the nucleation field ratio in the presence of the externally applied high frequency field : 
\begin{equation}
\frac{H_{c_3}^{new}}{H_{c_2}^{new}}=\frac{\langle H \rangle_{0,sosc.}^{wf}}{
\langle H \rangle_{0,dosc.}^{wf}}=\frac{0.4687}{0.2652}=1.7673.
\end{equation}
In view of Eq. (34) and Eq. (32), we can conclude that the nucleation field ratio for the driven system is slightly larger than the non-driven system. The basic inference that one can draw from the above discussion is that the high frequency field accentuates the surface critical nucleation field of superconductivity more than that of the bulk critical nucleation field.\\
\section{Summary \& Conclusions}
In this section, we briefly summarize all my derived results and then conclude this paper. We investigate the effect of high frequency electric field on the nucleation of superconductivity in the interior of a large sample as well as at the surface of a finite sample. Employing the linearized time dependent Ginzburg-Landau theory for this purpose we have shown the analogy between quantum single harmonic oscillator and the nucleation in the interior of a large sample. The similarities between the nucleation of superconductivity at the surface and a quantum double oscillator is also demonstrated. We invoke two approximate theories to derive critical nucleation field ratio $\frac{H_{c_3}}{H_{c_2}}$. The first approximate theory is based on the variational principle and is solved analytically. On the other hand the second approximate theory is based on the ground state energy method of A. P. van Gelder \cite{van}. We determine the minimum of the ground state energy by the numerical method. Since critical nucleation field is inversely proportional to the ground state energy one can easily compute the critical nucleation field ratio. The validity of these two approximated theories is checked by comparing the results obtained from these two theories with that of the exact results for the case of nondriven system and they agree very well with each other. The effect of high frequency oscillating time-periodic field is taken care of by the effctive time-independent Hamiltonian method of Cook et al \cite{cook}. It is observed that for electric field strength $\varepsilon = 0.01$ V/m and frequency $\omega=10$ GHz one can obtain 0.114 T and 0.189 T much more enhancement in the bulk nucleation and surface nucleation field respectively.\\
In conclusion, we examine in details the effect of high frequency field on the nucleation of superconductivity. Using the variational method and ground state energy method we have shown that the high frequency field actually accentuates the surface critical nucleation field of superconductivity more than the bulk critical nucleation field. The enhancement of the surface critical nucleation field is 1.6592 times more than the enhancement of the bulk critical nucleation field. One can now say that the ratio $\frac{H_{c_3}}{H_{c_2}}$ is not universal i.e. $H_{c_3}$ is not the universal upper limit of nucleation. One can obtain higher than $H_{c_3}$ field by applying high frequency oscillating electric field. Our results will be helpful in analyzing recently developed high-frequency superconducting devices \cite{asle,winkler}.\\  
{
\end{document}